\documentclass[10pt,a4paper]{article}
\usepackage[latin1]{inputenc}
\usepackage{amsmath}
\usepackage{amsfonts}
\usepackage{amssymb}
\usepackage{graphicx}
\begin{document}
\begin{center}
\large
\textbf{Parametrization Framework for the Deceleration Parameter in Scalar Field Dark Energy Model}\\
\end{center}
\begin{center}
A. Beesham\footnote{Email: abeesham@yahoo.com} \\
\vspace{2mm}
 \small 
 Faculty of Applied and Health Sciences, Mangosuthu University of Technology, Jacobs 4026,   South Africa\\
 
\end{center}
\vspace{3mm}
\textbf{Abstract:} We propose a Friedmann-Lemaitre-Robertson-Walker   cosmological model with a scalar field that represents  dark energy. A new parametrization of the deceleration parameter is introduced of the form \(q=-1+\eta/(1+\mu a^\eta)\) where \(\eta\) and \(\mu\) are model parameters. and the compatibility of the model is constrained by recent observational datasets, including cosmic chronometers,  Pantheon+ and Baryon Acoustic Observations. By considering a variable deceleration parameter, we address the expansion history of the  universe, providing a viable description of the transition from deceleration to acceleration. Using the Markov Chain Monte Carlo method, the parameters of the model are constrained and we examine the cosmological parameters. A comparison is then made with the  \(\Lambda\)CDM model using the latest  observations. We examine the history of the main cosmological parameters, such as the deceleration parameter, jerk parameter, snap parameter, density parameter, and equation-of-state parameter, by constraining and interpreting them to reveal insights into what has been dubbed ``dynamical dark energy" under the assumptions made above. Our method provides a  framework  that is independent of the model to explore dark energy, leading to a deeper and more subtle understanding of the mechanisms driving late-time cosmic acceleration.

\vspace{0.5cm}
\noindent
\textbf{Keywords}: Cosmological model; Dark energy; Current acceleration of  universe; Variable deceleration parameter;  Observational constraints

\vspace{0.5cm}
\noindent
\textbf{MSC code}: 83F05; 85A40

\vspace{0.5cm}
\noindent
\textbf{PACS}: 98.80.Jk; 95.30.Sf
	
\section{Introduction:}The accelerated expansion of the universe, first observed in the late 1990s through distant supernovae surveys, has revolutionized our understanding of the cosmos \cite{Reiss,2}. For an explanation of this, there are  two schools of thought by the researchers in the field. The first one is the introduction of a mysterious form of exotic matter called Dark Energy (DE) \cite{3,4,5}, which is characterized by  significant  pressure that is negative, and causes the current acceleration.  The second approach explores modified gravity theories by changing the gravitational action of general relativity \cite{6}. Some modified gravity theories which have been used to interpret the acceleration of the expanding universe include \( f(R) \) gravity \cite{7,8,9,10,11,12}, \( f(T) \) gravity \cite{13,14,15,16,17}, and \( f(Q) \) gravity, where functions of the Ricci scalar \( R \), torsion \( T \), and the non-metricity scalar \( Q \) are involved in modifying the gravitational action \cite{18,19,20,21,22,23}. These models can provide alternative explanations for the accelerated expansion of the universe due to changes in the expansion dynamics without invoking the need for exotic matter or dark energy. Besides these, there have been many Dark Energy (DE) models, such as quintessence \cite{24}, involving a dynamic scalar field; k-essence \cite{25}, which generalizes quintessence by considering more complex kinetic terms; phantom energy \cite{26,27,28}, characterized by an equation of state parameter less than -1; and scalar-tensor theories, which couple a scalar field with gravity. Each of these models introduces different theoretical frameworks and predictions, which are very helpful in gaining varying insights into the evolution of the universe, and its likely solutions to cosmological puzzles such as the coincidence problem and the nature of dark energy.

The \(\Lambda\)CDM model is extremely successful in describing a large variety of observational data, but has several theoretical challenges \cite{29,30,31}. One of the significant challenges is problem of fine tuning , arising from the  cosmological constant (\(\Lambda\)). It raises the question of fine tuning of  \(\Lambda\)  \cite{32}. The other major problem is problem of cosmic coincidence , which describes the peculiar fact that the dark energy (DE) and matter densities are aligned presently  \cite{33,34,35,36}. Additionally, new observations indicate that the \(\Lambda\)CDM model does not accurately describe the latest low-redshift cosmological data. Although the \(\Lambda\)CDM is consistent with many observations, some data suggest that the  DE density evolves over time, whereas it is constant in the standard model. These models are believed to fit observations better,

It is important to carefully consider and test all proposed models with the help of cosmological observations. This is because the mechanisms that drive late-time acceleration are complex and observational data is becoming increasingly precise. For such analyses, proper parametrizations need to be used that permit a description that is independent of the model for explaining the present \cite{37,38,39,40}.

In this paper, we discuss a Friedmann-Lemaitre-Robertson-Walker  model with a scalar field dark energy   and investigate its compatibility with recent observational datasets, including cosmic chronometers (CC), SNIa (Pantheon), and BAO. By considering a variable deceleration parameter (VDP), we address the  expansion history of the universe, providing a viable description of the transition from deceleration to acceleration. Using the Markov Chain Monte Carlo (MCMC) method, we constrain the model parameters and examine the cosmological quantities. In section 2, we provide the basics of the model, and relevant quantities. Then section 3 entails the determination of the model parameters using the MCMC (Markov Chain Monte CArlo) technique. In section 4,  we discuss our results, and finally, in section 5 is the conclusion.

\section{Cosmological model}
For the $k = 0$  Friedmann-Lemaître-Robertson-Walker (FLRW) universe, the metric is:
\begin{equation}
	ds^2 = dt^2 - a^2(t) \left[ dr^2 + r^2 (d\theta^2 + \sin^2 \theta d\phi^2) \right],\label{eq1}    
\end{equation}
where the symbols have their usual meanings.  We take the model to be populated by  two fluids that are perfect. One is  matter $m$, with negligible pressure (dark and baryonic matter), and the other is  a scalar field $\phi$. The latter, a scalar field, is  believed to cause  the present acceleration, and may be regarded as  Dark Energy (DE). In this scenario, Einstein's field equations and the Klein-Gordon equation for the scalar field can be written as follows (assuming \(8\pi G = c = 1\)):
\begin{equation}
	3H^2 = \rho_m + \rho_\phi = \rho_m + \frac{1}{2} \dot{\phi}^2 + V(\phi),\label{eq2}   
\end{equation}
\begin{equation}
	2\dot{H} + 3H^2 = -p_\phi = -\left(\frac{1}{2} \dot{\phi}^2 + V(\phi)\right),\label{eq3}    
\end{equation}
\begin{equation}
	\ddot{\phi} + 3H\dot{\phi} + \frac{dV}{d\phi} = 0.  \label{eq4} 
\end{equation}
Here, \( H = \frac{\dot{a}}{a} \) is the Hubble parameter, representing expansion or contraction.  The parameters \(p_\phi\), \(\rho_\phi\), and \(\rho_m\), are the pressure and  of the scalar field, and energy density  of matter,  respectively. The solution of the above equations with relevant initial conditions provides information about the evolution of the model. The evolution of $\phi$ with $m$  is given by: 
\begin{equation}
	\rho_m = \rho_{m0} a^{-3} = \rho_{m0} (1 + z)^3, \label{eq5}  
\end{equation}
where the subscript $0$ denotes the present time and since the redshift is $z= -1 +1/a(t)$. 

$\rho_\phi$   can be written  as the familiar equation:  $\rho_\phi = \frac{1}{2} \dot{\phi}^2 + V(\phi),$  
whereas $p_\phi$ is: $p_\phi = \frac{1}{2} \dot{\phi}^2 - V(\phi) $. As usual, the potential function is \( V(\phi) \). From equations (\ref{eq4}), (\ref{eq3}) and (\ref{eq2}),  We can get the evolution of the the Hubble parameter $H$, and \( V(\phi) \) in terms of $H$: 
\begin{equation}
	2\dot{H} = -\frac{\rho_{m0}}{a^3} - \dot{\phi}^2, \label{eq6}  
\end{equation}
and
\begin{equation}
	V(\phi) = + 3H^2 + \dot{H}  - \frac{\rho_{m0}}{2a^3}. \label{eq7} 
\end{equation}
From equation  (\ref{eq6}) we : $a \frac{d}{da} (H^2) + \frac{\rho_{m0}}{a^3} = -\dot{\phi}^2.$ 
In addition, we can express \( \dot{\phi} \) as $\dot{\phi} = aH \left( \frac{d\phi}{da} \right).$ 
The evolution of \( \phi \) in terms of $z$ is:
\begin{equation}
	\frac{d\phi}{dz} = \left[ (2E \frac{dE}{dz} - 3\Omega_{m0} (1+z)^2)\frac{1}{E^2(1+z)} \right]^{1/2},  \label{eq8}
\end{equation}
where  represents the dimensionless  \( E(z) = \frac{H(z)}{H_0} \) is the Hubble parameter without dimension, and \( \Omega_{m0} = \frac{\rho_{m0}}{3H_0^2} \)is the current  density  of matter.
Similarly, $V(z)$ is given by:
\begin{equation}
	\frac{V(z)}{3H_0^2} = - (1 + z)^3E \frac{dE}{dz} + E^2 - \frac{1}{2} \Omega_{m0} (1 + z)^3.\label{eq9}  
\end{equation}

The definition of the deceleration parameter  is:
\begin{equation}
	q(z) = - \frac{\dot{H}}{H^2} - 1 = \frac{dE}{dz} \frac{(1 + z)}{E}  - 1.  \label{eq10}
\end{equation}
Deceleration is indicated by  \( q > 0 \), acceleration by \( q < 0 \) and a constant expansion by \( q = 0 \).   
For \( q = -1 \), we get de Sitter expansion  (exponential expansion) , and finally,    \( q < -1 \) denotes expansion that is super-exponential.

\( \Omega_m \) and \( \Omega_\phi \) in terms of the redshift $z$ are given by:
\begin{equation}
	\Omega_m(z) = \frac{\rho_m}{3H^2} = \frac{\Omega_{m0} (1 + z)^3}{E^2}, \label{eq11}  
\end{equation}
\begin{equation}
	\Omega_\phi(z) = 1 - \Omega_m(z) = 1 - \frac{\Omega_{m0} (1 + z)^3}{E^2}. \label{eq12}  
\end{equation}
Another parameter useful in the study of dark energy is the  equation of state  parameter (eos) denoted by  \( \omega_\phi(z) \), defined by:
\begin{equation}
	\omega_\phi(z) = \frac{p_\phi}{\rho_\phi} =\frac{-1 - \frac{2\dot{H}}{3H^2}}{\Omega_\phi}  . \label{eq13}  
\end{equation}
Thus, we get:
\begin{equation}
	\omega_\phi(z) =\frac{\frac{2}{3} (1 + z) E \frac{dE}{dz} - E^2}{E^2 - \Omega_{m0} (1 + z)^3} . \label{eq14}   
\end{equation}

\section{Kinematical Parameters}
We now choose \( q \) as :
\begin{equation}
	q = -1+\frac{\eta}{1+\mu a^\eta}. \label{eq15}
\end{equation}
for which
\begin{equation}
	H=\kappa (\mu+ a^{-\eta}),\label{eq16}
\end{equation}
The motivation for this form of $q$ comes from a recent study by Pawde et al \cite{41} who considered $q = -1+\eta/(1+a^\eta)$. As no observational constraints were provided via an MCMC analysis, we made this study \cite{42}, and found that that form for $q$ did not fit the observations. Hence, in this investigation, we consider form (\ref{eq15}). Another obvious motivation for the condition (\ref{eq15}) is as follows:
For large scale factor $a$, we have  $q \rightarrow -1$, as is the case for the  $\Lambda$CDM model. Hence the assumption of this condition ensures that the model will approach the $\Lambda$CDM model in future. 

Additional  motivation for our choice of deceleration parameter $q$ as in Eq. (\ref{eq15}) is as follows. 
\begin{itemize}
	\item The study of cosmological models within the climate of  late time acceleration is expressed in terms of kinematic  parameters such as $q$. 
	\item Expressing $q$ as $q = -(\ddot{a}/a)/(H^2) $, we see that it is essentially the acceleration divided by the expansion. In the past, the universe was decelerating, so $q > 0$, or $q = const > 0$. However, now we are experiencing acceleration, so $q<0$. Hence, for continuity, we need a dynamic $q$ as a function of time (or redshift), that changes sign from positive to negative.
	\item Hence, many form of $q$ have been adopted to try to explain this transition \cite{43,44} (and references therein).
	\item Since $a=1/(1+z)$, for our choice of $q$, we can make some qualitative remarks about how $q$ varies with redshift. 
\end{itemize}

\begin{itemize}
	\item Both parameters $\eta$ and $\mu$ relate to $q_0$, i.e., $q_0 = -1 + \eta/(\mu + 1)$. Presently,  the state of the model  depends upon the value of these parameters.  $\mu = \eta - 1 \implies  q_0 = 0$, and the universe is undergoing constant expansion.  $\mu > \eta - 1 \implies q_0 > 0$, and we have decelerated expansion. Finally, if $\mu < \eta - 1$, then $q_0 < 0$, and we have accelerated expansion. 
	\item In the distant past, $z>>1$, and $q(z) \rightarrow -1 + \eta$. For $\eta > 1$, we have $q>0 \implies$ deceleration, i.e., the radiation and matter dominated eras.
	\item In the far future, from the form of $q(a)$ as in Eq. (\ref{eq15}), and the discussion following that equation, we have seen that $q  \rightarrow -1$, the asymptotic form for $q$ for the $\Lambda$CDM model. Hence this model will asymptotically approach the $\Lambda$CDM model in future.
\end{itemize} 

We can write in terms of redshifts
\begin{equation}
	H=\frac{H_0}{1+\mu}[\mu+ (1+z)^{\eta}].\label{eq17}
\end{equation}
Consider the special form (\ref{eq15}) of the deceleration parameter; Equations (\ref{eq14}), (\ref{eq12}), (\ref{eq11}) and (\ref{eq10}), and  can be written as:
\begin{equation}
	q=-1+\frac{\eta (1+z)^\eta}{\mu +(1+z)^\eta},\label{eq18}
\end{equation}
\begin{equation}
	\Omega_m (z)=\frac{\Omega_{m0}(1+\mu)^2(1+z)^3}{{[\mu +(1+z)^\eta]^2}},\label{eq19}
\end{equation}
\begin{equation}
	\Omega_\phi (z)=1-\frac{\Omega_{m0}(1+\mu)^2(1+z)^3}{{[\mu +(1+z)^\eta]^2}},\label{eq20}
\end{equation}
\begin{equation}
	\omega_\phi (z)=\frac{\frac{2\eta}{3}(1+z)^\eta [\mu +(1+z)^\eta] -[\mu +(1+z)^\eta]^2}{[\mu +(1+z)^\eta]^2 -\Omega_{m0}(1+\mu)^2 (1+z)^3}.\label{eq21}
\end{equation}
The jerk and snap parameters of cosmology give additional higher derivatives of the scale factor for the universe as opposed to the traditional two  parameters (Hubble and deceleration). These higher order derivates  describe the cosmological expansion in  further and finer detail.
\begin{equation}
	j = \frac{1}{a} \frac{d^3 a}{d\tau^3} \left(\frac{1}{a} \frac{da}{d\tau}\right)^{-3} = q(2q + 1) + (1 + z)\frac{dq}{dz},
\end{equation}
\begin{equation}
	j(z)=\left( -1+\frac{\eta (1+z)^\eta}{\mu +(1+z)^\eta}\right)\left( 1+\frac{2\eta (1+z)^\eta}{\mu +(1+z)^\eta}\right)+\frac{\eta^2 \mu (1+z)^\eta}{[\mu +(1+z)^\eta]^2},\label{eq22}
\end{equation}
\begin{equation}\nonumber
	s= \frac{j-1}{3(q-\frac{1}{2})} , 
\end{equation} 
\begin{equation}\nonumber
	s= \frac{\left( -1+\frac{\eta (1+z)^\eta}{\mu +(1+z)^\eta}\right)\left( 1+\frac{2\eta (1+z)^\eta}{\mu +(1+z)^\eta}\right)+\frac{\eta^2 \mu (1+z)^\eta}{[\mu +(1+z)^\eta]^2}-1}{3\left(-\frac{3}{2}+\frac{\eta (1+z)^\eta}{\mu +(1+z)^\eta}\right)}.\label{eq24} 
\end{equation} 

\section{Determination of Model Parameters Using MCMC Technique}

In cosmology, Bayesian methods are commonly used to estimate parameters by computing the posterior distribution of the parameters \( \theta \) based on observed data \( D \):
\begin{equation}
	P(\theta \mid D) = \frac{L(D \mid \theta) P(\theta)}{P(D)},   
\end{equation}
with \( P(\theta) \) being the prior distribution, \( P(D) \)  the marginal likelihood and \( L(D \mid \theta) \)  the likelihood function. Bayesian parameter estimation involves exploring the parameter space \( \theta \), often with algorithms like Metropolis-Hastings \cite{45}, which helps guide a random walker through the space, preferring regions with higher likelihoods. The mean and uncertainty of each parameter are usually found by analyzing where the walker spends  most of its time and how far it deviates within the parameter space. In situations where we have a nearly Gaussian posterior distribution, information criteria offers a simpler approach to model selection \cite{46}. In this work, we investigate the form (\ref{eq18}) for the deceleration parameter.  A Markov Chain Monte Carlo (MCMC) analysis  with the emcee package \cite{47} is used to properly cover the   parameter space in order to get   reliable  estimates. The GetDist package \cite{48} is employed to visualize  and plot the posterior distributions. This enables the proper determination of contraints for the parameters. 

\subsection{ Cosmic chronometers (CC)}

CC are strong probes of cosmic expansion, and offer a model-independent way to estimate the Hubble parameter \( H(z) \). We can calculate  \( H(z) \) at a given redshift \( z \) from the metallicity and age of passive nearby  galaxies. This approximation  comes from the formula: 
$H(z) \approx - (\Delta z / \Delta t)/(1 + z)$. CC data is obtained from a number of sources \cite{49,50,51,52,53,54}, over the redshift range \( 0.07 \lesssim z \lesssim 1.97 \) \cite{46}. This data provides the  constraints on $H(z)$.  To determine how well theoretical models fit in with CC  data at any given redshift, we calculate the chi-squared statistic \( \chi^2 \):
\begin{equation}
	\chi^2_{\mathrm{CC}}(\theta) = \Delta H^{T}(z) C^{-1} \Delta H(z), \end{equation}
where \( \Delta H(z) \) is the difference between the predicted expansion rate of the model, \( H_{\mathrm{M}}(z) \), and that of cosmic chronometer data, \( H_{\mathrm{D}}(z) \), at a given redshift $z$, and \( C \) the covariance matrix.

\subsection{ Type Ia supernova (SNIa)}

The Pantheon+ dataset consists of light curves for 1701 Type Ia Supernovae (SNe Ia) from 1550 unique events with redshifts in the range \( 0 \leq z \leq 2.3 \) \cite{55}. The  apparent magnitude of such a star is:
\begin{equation}\label{appmagn}
	m(z) = 5 \log_{10} \left( \frac{d_L(z)}{\text{Mpc}} \right) +\mathcal{M} + 25,    
\end{equation}
where  \( \mathcal{M} \) is the absolute magnitude of the star and $d_L(z)$  its luminosity distance given by:
\begin{equation}
	d_L(z) = (1 + z) \int_0^z \frac{dz'}{H(z')H_0},   
\end{equation}
For the SNe Ia data, 
\begin{equation}
	\chi^2_{\mathrm{S}} = \Delta D^T C_{\mathrm{t}}^{-1} \Delta D,   
\end{equation}
where the total covariance matrix \( C_{\mathrm{t}} = C_{\mathrm{sys}} + C_{\mathrm{stat}} \) is the sum of the systematic and statistical covariance matrices, respectively. The  deviation of the   distance modulus of a star  is given by:
\begin{equation}
	\Delta D = \mu(z_i) - \mu_{\mathrm{model}}(z_i, \theta),   
\end{equation}
where \( \mu(z_i) = m(z_i) - \mathcal{M} \) is the observed distance modulus. We project \( \mathcal{M} \) up to a normalization constant in the likelihood function \( L \propto e^{-\chi^2 / 2} \) \cite{56}.

\subsection{ Baryon acoustic oscillations (BAO)}
The sound horizon, \( r_d \), defined at the epoch of baryon decoupling (\( z_d \approx 1060 \)), is given by:
\begin{equation}
	r_d = \frac{1}{H_0} \int_{z_d}^{\infty} \frac{c_s(z)}{E(z)} dz,   
\end{equation}
where \( c_s(z) \) is the sound speed, a function of the baryon-to-photon density ratio, and \( E(z) \equiv H(z)/H_0 \) is the dimensionless Hubble parameter. Now, in cosmology, measurements of Baryon Acoustic Oscillations (BAO) depend on \( r_d \).  In this work, we take  \( r_d \) as a free parameter \cite{57,58,59,60}, rather than as a prior based on CMB Planck data .

\begin{figure}  
	\centering  
	\includegraphics[width=1.0\textwidth]{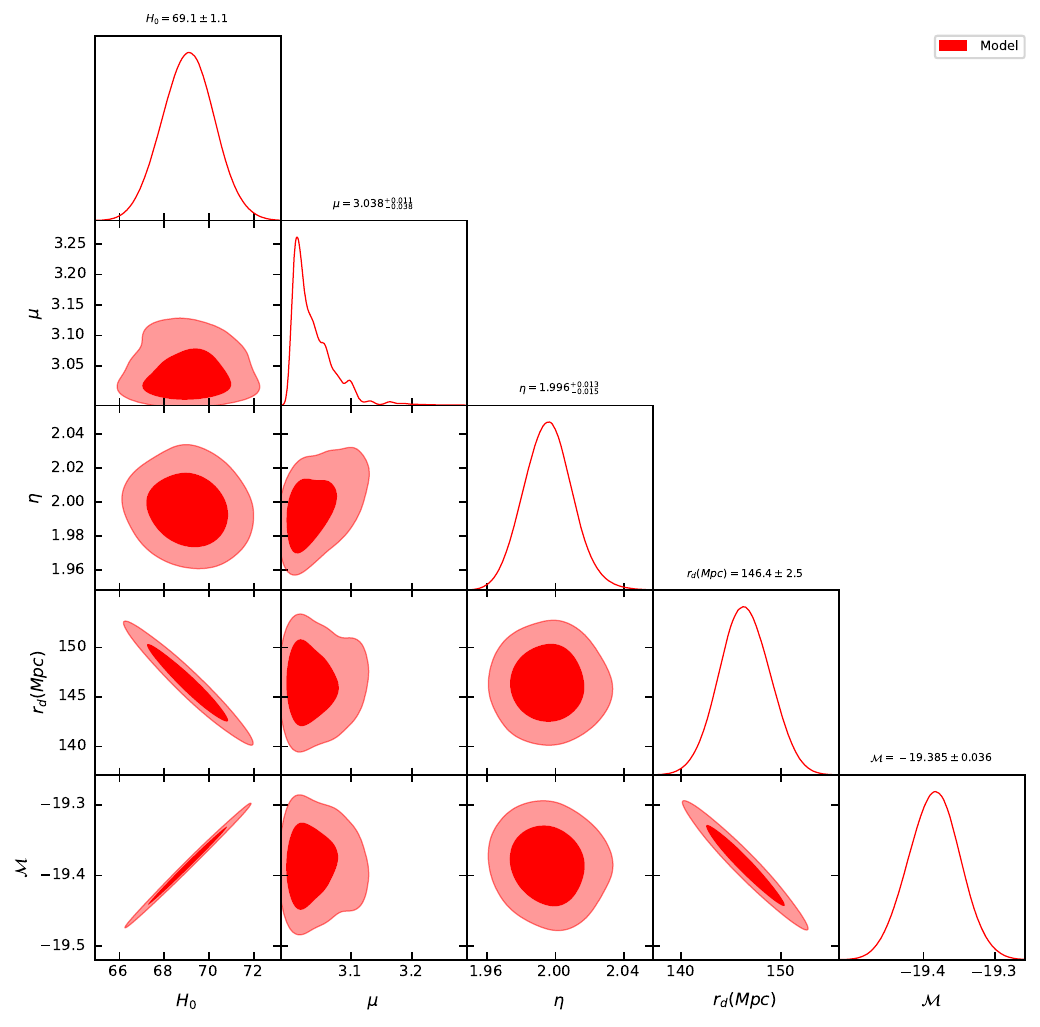} % Name and format of the figure file  
	\caption{Posterior distributions of parameters for  cubic varying deceleration parameter}  
	\label{fig:mcmc_curve}  
\end{figure}

Data from the completed Sloan Digital Sky Survey (SDSS-IV) \cite{61} and the BAO catalogs from the first-year observations of the Dark Energy Spectroscopic Instrument (DESI Y1) \cite{62} are utilised in the analysis. The distance measures that are used are the Hubble distance  \( D_H(z)= c/H(z)\), the comoving angular diameter distance \( D_M(z)/r_d \):
\begin{equation}
	D_M(z) = \frac{c}{H_0} \int_0^z \frac{dx'}{E(x')}.   
\end{equation}
and 
the volume-averaged distance \( D_V(z)/r_d \), which encodes the position of the BAO peak:
\begin{equation}
	D_V(z) = \left[ z D_M^2(z) D_H(z)  \right]^{1/3}.  
\end{equation}
The \( \chi^2 \) statistic for the distance measurements scaled by the sound horizon, \( D_X/r_d \), is:
\begin{equation}
	\chi^2_{D_X/r_d} = \Delta D_X^T C_{D_X}^{-1} \Delta D_X,    
\end{equation}
where \( \Delta D_X = D_{X, \mathrm{Model}}/r_d - D_{X, \mathrm{Data}}/r_d \) for \( X = H, M, V \), and \( C_{D_X}^{-1} \) is the inverse covariance matrix for each \( X \). In Figure \ref{fig:mcmc_curve}, we have plotted the confidence contours for each parameter of our choice  of the varying deceleration parameter (\ref{eq18}). Table I contains the best-fit values for the parameters of our model.

\begin{table*}
	\begin{center}
		\begin{tabular}{|c|c|c|c|}
			\hline
			\multicolumn{4}{|c|}{MCMC Results} \\  
			\hline
			Model & Parameters  & Prior & Joint   \\[1ex]
			\hline
			& $H_0$ &[50,100] & $69.1\pm 1.1$  \\[1ex] 
			&$\mu$  & [3,4] & $3.038 ^{+0.011}_{-0.038}\pm 0.031$ \\[1ex]
			VDP model&$\eta$& [1,2]& $1.996^{+0.013}_{-0.015}$   \\[1ex]
			&$\mathcal{M}$ & [-20,-18] & $-19.385\pm 0.036$ \\[1ex]
			&$r_d (Mpc)$ & [100,200] & $146.4\pm 2.5$  \\[1ex]   
			\hline
		\end{tabular}
		\caption{The calculated best-fit values for model using  \texttt{CC+SNIa+BAO} datasets.}
		\label{tab_MCMC}
	\end{center}
\end{table*}

\begin{figure}  
	\centering  
	\includegraphics[width=0.9\textwidth]{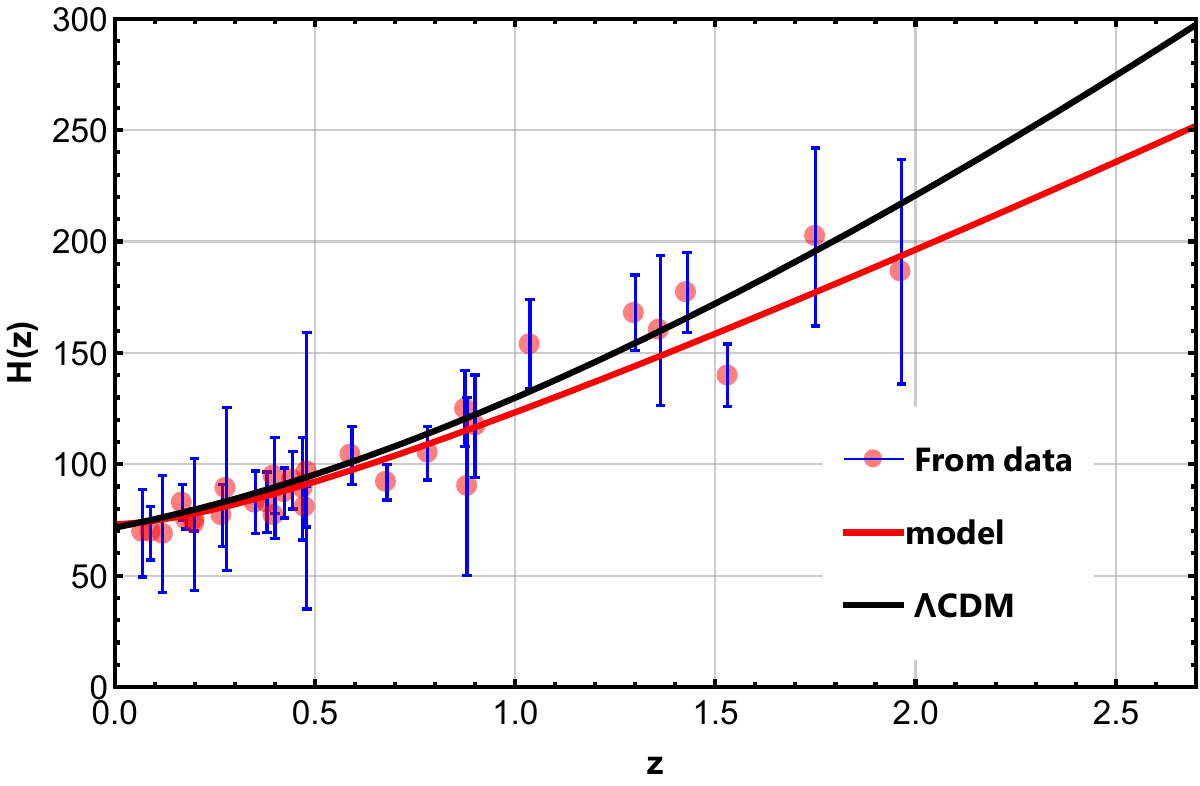} % Name and format of the figure file  
	\caption{Comparison of the Hubble parameter curve with the \(\Lambda\)CDM model.}  
	\label{fig:Hubble_curve}  
\end{figure}

\begin{figure} 
	\centering  
	\includegraphics[width=0.9\textwidth]{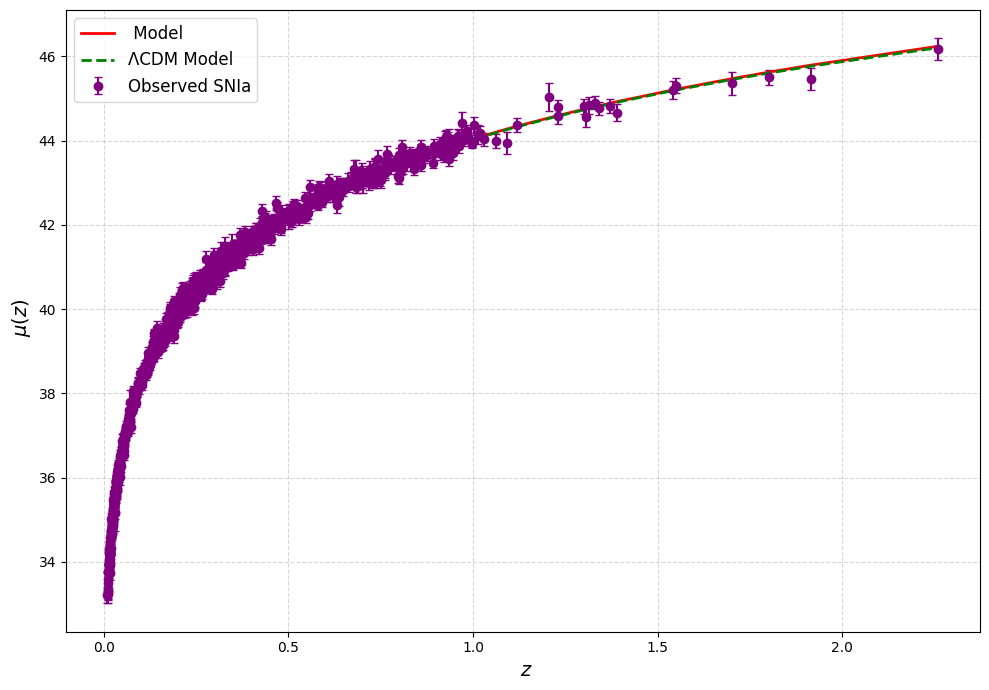} % Name and format of the figure file  
	\caption{Comparison of the distance modulus curve with the \(\Lambda\)CDM model.}  
	\label{fig:mu_curve}  
\end{figure}

Additionally, Fig.~\ref{fig:Hubble_curve} compares the Hubble parameter curve predicted by the model with the \(\Lambda\)CDM model, demonstrating consistency with the observational data, and Fig. \ref{fig:mu_curve} gives a comparison of the distance modulus curve with the $\Lambda$CDM model.

\section{Results and discussion}
\begin{figure} 
	\centering  
	\includegraphics[width=0.9\textwidth]{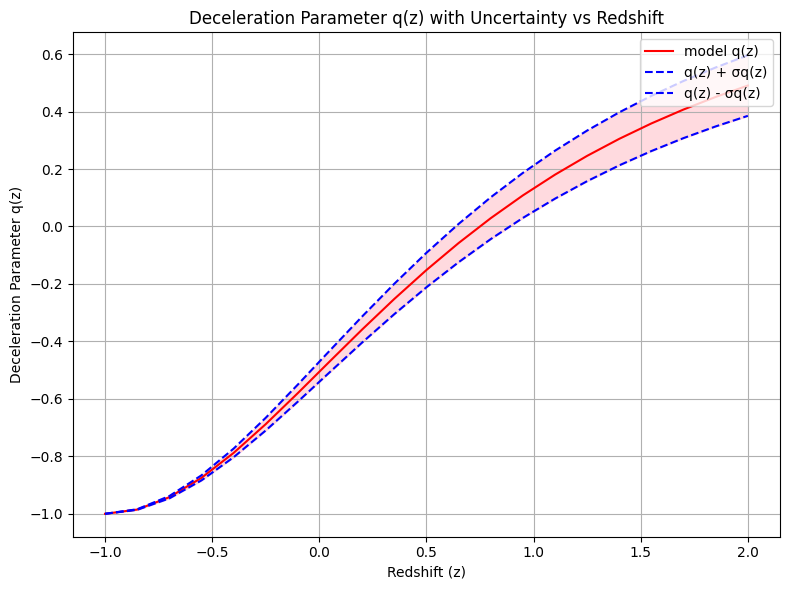} % Name and format of the figure file  
	\caption{ \( q(z) \) vs \( z \) with $1 \sigma$ error bounds.}  
	\label{fig:q_curve}  
\end{figure}

We have already obtained the best fit values of the parameters  $H_0$, $\eta$ and $\mu$  in section 4. Next, we discuss the cosmological evolution of the model  variables as constrained by the observations. Our primary focus was on the  combined data sets: $\mathcal{CC+SNIa+BAO}$. Based on the best-fit values for $H_0$, $\eta$ and $\mu$  from Table 1, we continue our evaluation in this  section. $q$ was reconstructed,   and this is  illustrated   in Fig.~\ref{fig:q_curve}, together with the 1 $\sigma$ error bounds. The figure   illustrates that at a best-fit transition redshift of \(z_{\text{tr}} = 0.748^{+0.4}_{-0.4}\), \( q(z) \) change sign from positive to negative. This means a transition from deceleration (required for structure formation, etc) to acceleration.  The current value of \( q(z) \) is  
\(q_0 = -0.57^{+0.46}_{-0.46}, \). $z_{tr}$ and $q_0$ are within observational constraints. 
\cite{63,64}.

\begin{figure}  
	\centering  
	\includegraphics[width=0.9\textwidth]{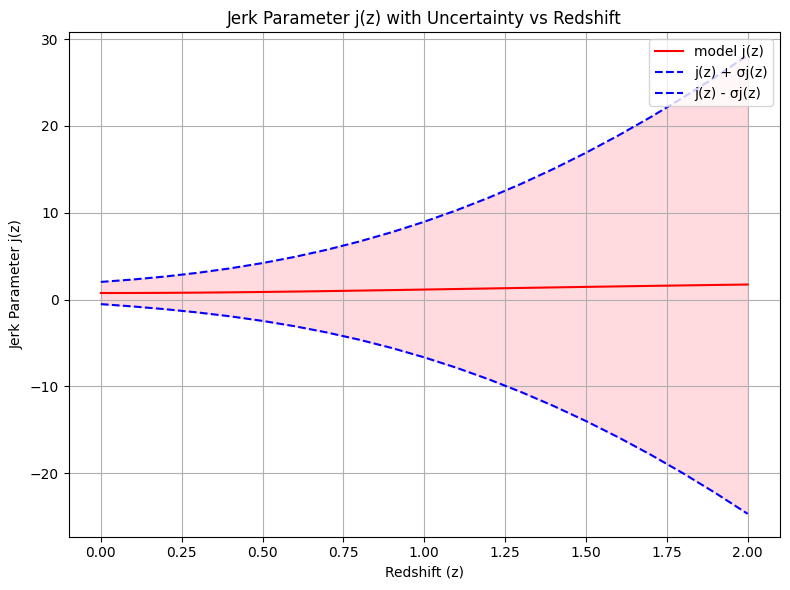} % Name and format of the figure file  
	\caption{Jerk parameter \( j(z) \) vs redshift \( z \) with $1 \sigma$ error bounds.}  
	\label{fig:j_curve}  
\end{figure}

The jerk parameter \cite{65} is significant in cosmology, as it is essentially the the third order term in the Taylor series expansion of $a(t)$.  It is a unique characterization of cosmic dynamics.  It  contains useful information related to the evolution of the cosmos and distinguishes between various dark energy models. It acts as an indispensable bridge between dark energy and normal cosmological models. The different values of \( j \) establish a relationship between several theories of dark energy and the  \( \Lambda \)CDM model; for instance, \( j = 1 \) refers to the  \( \Lambda \)CDM model. A grasp of the jerk parameter is fundamental to study the dynamics in cosmic expansion and the transitions between different eras of acceleration. The jerk parameter for our model is sketched in Fig. \ref{fig:j_curve}, and it may be seen, surprisingly, that $j$ is constant, with $j=1$. This result has been obtained by starting with a form of the deceleration parameter that was motivated by  by Pawde et al \cite{41}.

\begin{figure}  
	\centering  
	\includegraphics[width=0.9\textwidth]{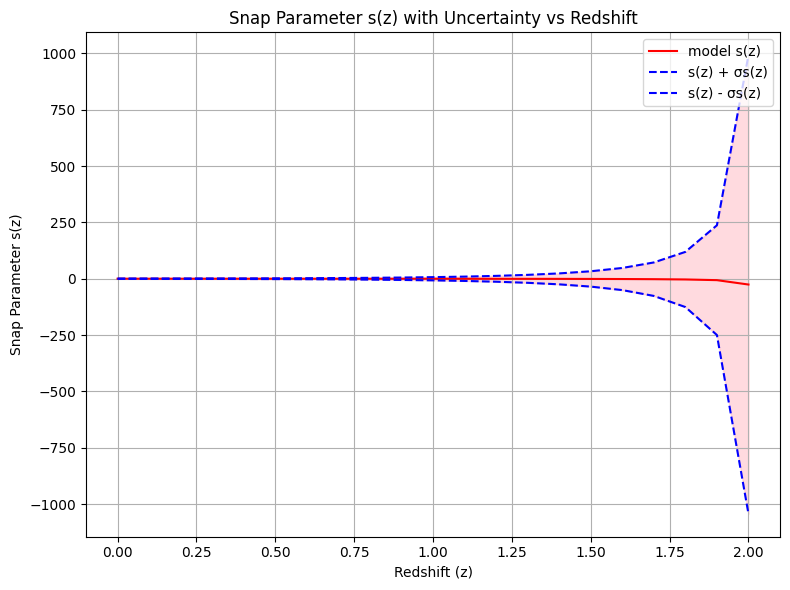} % Name and format of the figure file  
	\caption{ \( s(z) \) versus \( z \) }  
	\label{fig:s_curve}  
\end{figure}

\begin{figure} 
	\centering  
	\includegraphics[width=0.9\textwidth]{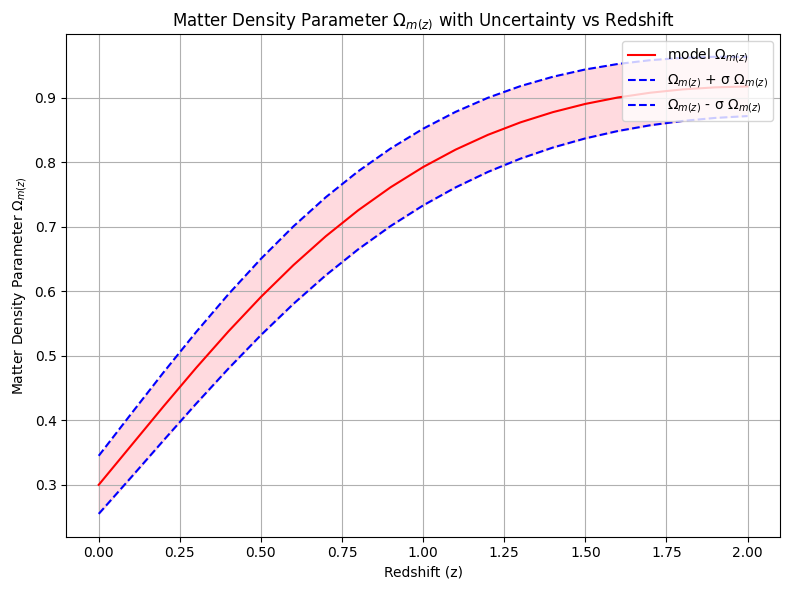} % Name and format of the figure file  
	\caption{ \( \Omega_m \) vs \( z \). }  
	\label{fig:m_curve}  
\end{figure}

\begin{figure}  
	\centering  
	\includegraphics[width=0.9\textwidth]{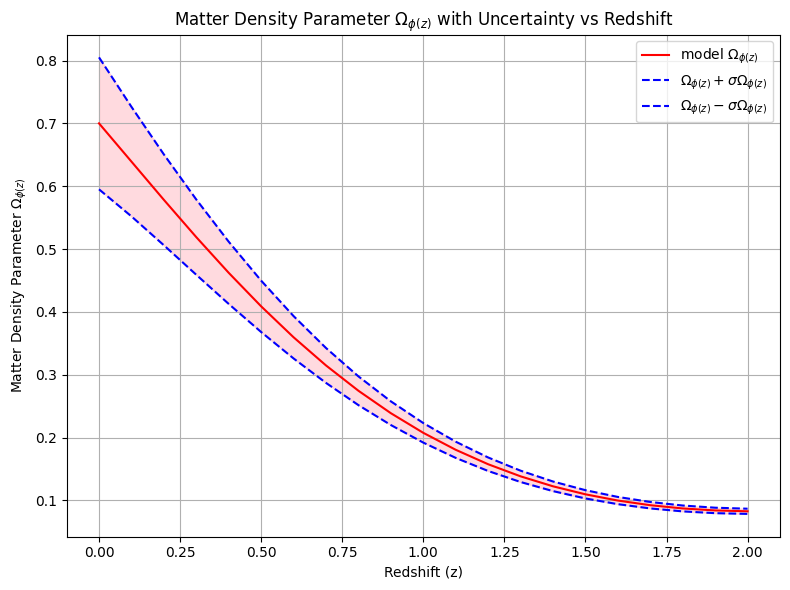} % Name and format of the figure file  
	\caption{ \( \Omega_\phi \) versus  \( z \).}  
	\label{fig:m1_curve}  
\end{figure}

\begin{figure}  
	\centering  
	\includegraphics[width=0.9\textwidth]{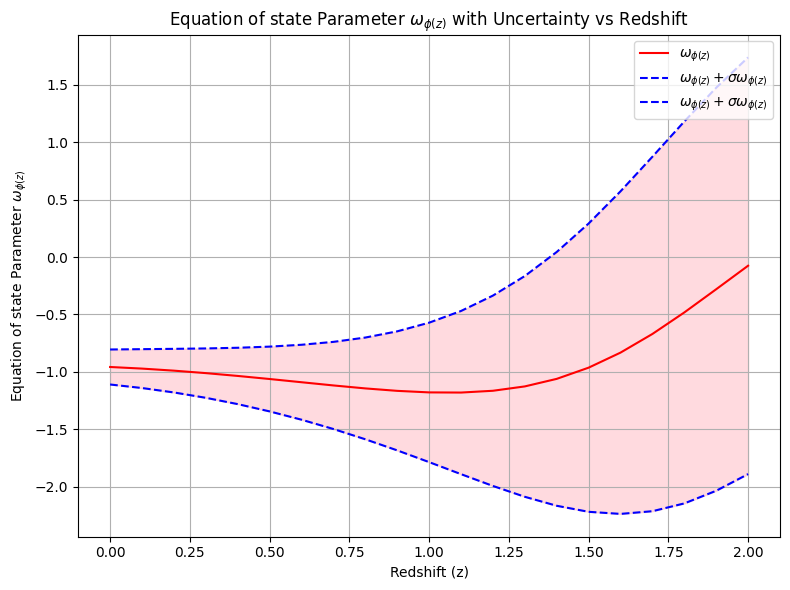} % Name and format of the figure file  
	\caption{ \( \omega_\phi \) vs \( z \).}  
	\label{fig:w_curve}  
\end{figure}

The snap parameter \cite{65}, which is represented by \( s \), is a cosmological parameter that gives the fourth time derivative of the scale factor and hence provides insight into how the curvature and expansion dynamics of the universe are set. It plays an important role in the Taylor series expansion that describes the growth of the Universe.  \( j = 1 \) for  the \( \Lambda \)CDM model,  and so we get \( s = -(2 + 3q) \).  The snap parameter for our model is plotted in Fig. \ref{fig:s_curve}, from which it can be seen that it is constant, $s = 0$.
Figs.~\ref{fig:m_curve} and Fig.~\ref{fig:m1_curve} are the plots  of the density parameters for $m$ and $\phi$, respectively. During early times, $m$  dominates, while $\phi$ is negligible.  With time,  the influence of $m$ decreases, whilst that of $\phi$ increases. matter density parameter decreases due to expansion. However, in the course of time, the scalar field density parameter becomes dominant and overshoots that of matter. This leads to acceleration of the expansion of the universe, which is the critical point in cosmic evolution. Furthermore, the density parameters at present time have been determined to be \(\Omega_{m0}=0.33 ,  \quad \Omega_{\phi 0} =0.7\)
for the $\mathcal{CC+SNIa +BAO}$ datasets. In the \( \Lambda \)CDM model, \(\Omega_{m0} = 0.315 \pm 0.007\)  \cite{66,67,68}, and  the values for our model is in agreement with those of the \( \Lambda \)CDM model.
Fig.~\ref{fig:w_curve} shows  \( \omega_\phi \) versus  $z$. It starts from the quintessence region, moves into the phantom phase for \( 0.35< z< 1.45\), and then back into the quintessence region (\( \omega_\phi > -1 \)).   Furthermore,  \( \omega_{\phi 0} = -0.99\), in excellent agreement with observations. 
\cite{69,70}.

\section{Conclusion}
In this paper, we investigated a scalar field dark energy model with a specific form (\ref{eq15}) of the deceleration parameter. The motivation for this form was a simpler form studied by   Pawde et al \cite{41} who considered $q = -1+\eta/(1+a^\eta)$. This was  without the parameter $\mu$ as compared to our choice.  We found that that choice by Pawde et al was not compatible with observations \cite{42}.  We began by giving a background to the dark energy cosmological, together with all the relevant equations. We considered baryonic and dark matter together with a scalar filed which represents dark energy. Then we motivated for our choice of deceleration parameter, and gave all the kinematic parameters in terms of z, such as the Hubble, jerk and snap parameters. In addition, we gave the energy parameters for the matter and scalar field. 

Then in section 3, we subjected our model to observational constraints te determine the parameters of our model. Using a combination of cosmic chronometers, Pantheon and baryon acoustic oscillation datasets, we found the following parameters: the present value of the Hubble parameter $H_0$, the constants $\mu$  and $\nu$ in our form (\ref{eq15}) of the deceleration parameter, along with their corresponding 1-\(\sigma\) and 2-\(\sigma\) confidence regions, which are given in Table 1 and Fig.~\ref{fig:mcmc_curve}, respectively. The  absolute magnitude $M$ and the sound horizon $r_d$ were also constrained by observations. We then compared our Hubble parameter $H(z)$ and distance modulus curve with that of the $\Lambda$CDM model, finding a good fit.

We then plotted the cosmographic parameters $q(z)$, $j(z)$ and $s(z)$. The current value $q_0 = -0.57$ of the deceleration parameter, and transition redshift $z_{tr} = 0.748$ are well within observational constraints, and the jerk and snap parameters, surprisingly, are the same as that of the $\Lambda$CDM model. Thus our model is viable, and provides an alternative to the standard $\Lambda$CDM model. In view of the importance of scalar fields, we feel that this model has the potential to contribute to the knowledge of dark energy, and towards a better understanding of the current acceleration of the universe.

\end{document}